\documentclass{article}

\usepackage[english]{babel}

\usepackage[letterpaper,top=2cm,bottom=2cm,left=3cm,right=3cm,marginparwidth=1.75cm]{geometry}

\usepackage{amsmath}
\usepackage{graphicx}
\usepackage{authblk}
\usepackage[colorlinks=true, allcolors=blue]{hyperref}

\title{Triaxiality dynamics of quadrupole excitation in even-even heavy nuclei}
\author{M. S. Nadirbekov\footnote{mnadirbekov@yandex.ru}, O.A. Bozarov, S.N. Kudiratov}
\affil{Institute of Nuclear Physics, Tashkent, 100214, Uzbekistan}

\begin{document}
\maketitle

\begin{abstract}
Abstract:
In the framework free triaxiality model branching ratio reduced E2-transitions probabilities are studied for the full region changes of the triaxiality parameter. The variables of quadrupole deformations $\beta_2$ and $\gamma$ are dynamic. The Davidson potential for $\beta_2$ and $\gamma$ variables was being used. In the free triaxiality model the energy levels was being obtained by taking into account the high-order terms of the rotational energy operator series, which leads energy levels of ground-state-band, $\beta$- and $\gamma$-bands to improve considerably the agreement of the results with experimental data. The quadrupole moment of the nuc\-leus expressed in terms of two internal quadrupole moments: $Q_0$ is quadrupole moment about the symmetry axis, $Q_2$ is the measure of the shape asymmetry about the symmetry axis. The intra/inter-band reduced E2-transition probabilities ground-state-band, $\beta$- and $\gamma$-bands are being presented and are being expressed in terms of four parameters: $\mu_{\beta_2}$, $\mu_\gamma$, $\gamma_0$ and $Q_0$. The influence of $\gamma$-oscillations to the reduced probabilities of E2-transitions takes into account  are numerically. The sensitivity of the intra-/inter-band reduced E2-transition probabilities to the triaxiality parameter was analyzed.

\end{abstract}

\section{Introduction}

\hspace{0.6 cm} Most of the atomic nuclei are deformed in their
ground states and possess axially symmetric prolate or oblate shapes
\cite{bohr,wil,iach}. It is known that the prolate shapes
essentially dominate over the oblate ones \cite{bonat1}. However,
various theoretical studies suggest for some nuclear regions the
possible deformations of each nuclide have been examined. The possible appearance of non-axial (triaxial) deformations has been
suggested \cite{bohr,davydov,davyd,chaban,max,hirata,monica,1,dav5,por,pnu}.

Quadrupole shapes are described by the parameters $\beta_2$ and
$\gamma$ for the axial deformation and the deviation from axiality
\cite{davydov,davyd,chaban,max,por}. Especially the majority of deformed nuclei correspond to an axially symmetric, elongated shape ($\gamma$=0$^0$) \cite{bohr}. Oblate forms are less common, and few nuclei are expected to be oblate in their ground state ($\gamma$=60$^0$) \cite{bohr}. Intermediate values of $\gamma$, i.e. 0$<\gamma<\pi$/3, give triaxial shapes. Bohr and Mottelson \cite{bohr} noticed that at $\gamma\geq$24$^0$ the nuclei do not remain deformed in the original sense and the nucleus can take on any shape, including triaxial one.

In Ref.\cite{davyd} the free triaxiality model was
proposed, where deformation parameters
$\beta_2$ and $\gamma$ taking into account are dynamic. In Ref.\cite{max}, a comparative analysis of some
non-adiabatic models \cite{davyd,chaban,min,tar} was carried out,
and it was determined that the free triaxiality model reproduces the
experiment better than other models.  In
Ref.\cite{por}, the free triaxiality model was developed taking into
account the high-order terms of the rotational energy operator
series. In Ref.\cite{nadir1} taking into account the high-order terms of the rotational energy operator on variable $\gamma$  leads to improve considerably the agreement of
the results with experimental data. And in Ref.\cite{nadir2} intra-/inter-band reduced \emph{E}2-transitions probabilities in excited
collective states of  even-even lanthanide and actinide nuclei was
studied. In Ref.\cite{gupta} one-parameter Davydov-Filippov model has been used to study the intra-/inter-bands transition B(E2) ratios  in ground and $\gamma$-bands for the triaxially deformed nuclei. But in this work low-spin states are restricted and variables $\beta_2$ and $\gamma$ are statistic.

Thus, in the works cited above, deformed triaxial nuclei were
studied, but the full range of variation of the deformation parameter $\gamma$ ($0<\gamma<\pi/6$) was not taken into account. Present work we will solve this problem. The  intra-/inter-band reduced E2-transitions probabilities we consider by taken into account high spin states. These data with the properties of the energy spectrum can provide information about the nature excited collective states of the even-even heavy nuclei including high spin states. 

\section{Free triaxiality approximation}
In Ref. \cite{max} we consider the possibility of describing the
energy levels of the ground-state-band, $\gamma$-rotational, and
$\gamma$- and $\beta$-rotational-vibrational bands by the Hamilton
operator \cite{davyd} containing five dynamic variables:
\begin{equation}
\hat{H}_{\beta_2}=\hat{T}_{\beta_2}+\hat{T}_{\gamma}
+\hat{T}_{rot}+V(\beta_2,\gamma), \label{quadrupol}
\end{equation}
here
\begin{equation}
\hat{T}_{\beta_2}=-\frac{\hbar^2}{2B_2}\frac{1}{\beta^4_2}\frac{\partial}{\partial\beta_2}
\left(\beta^4_2\frac{\partial}{\partial\beta_2}\right),
\label{beta2}
\end{equation}
\begin{equation}
\hat{T}_{\gamma}=-\frac{\hbar^2}{2B_2}\frac{1}{\beta^2_2\sin(3\gamma)}\frac{\partial}{\partial\gamma}
\left[\sin(3\gamma)\frac{\partial}{\partial\gamma}\right],
\label{gamma2}
\end{equation}
\begin{equation}
\hat{T}_{rot}=\frac{1}{4}\sum^{3}_{\lambda=1}\frac{\hat{I}^2_{\lambda}}{\left[sin\left(\gamma-\frac{2\pi}{3}\lambda\right)\right]^2},
\label{rot}
\end{equation}
$V(\beta_2,\gamma)$ -- potential energy of $\beta_2$- and $\gamma$-vibrations.

It is known that simple special solutions of the Bohr Hamiltonian,
which come from the exact separation of variables in the
corresponding Schr$\ddot{o}$dinger equation \cite{wil}, can be
obtained when the potential $V(\beta_2,\gamma)$ is represented as $
V(\beta_2,\gamma)=V(\beta_2)+V(\gamma)$. The solution of the
Schr$\ddot{o}$dinger equation for this form of potential was
obtained for the value of the $\gamma$-variable $\gamma\approx0$
\cite{iach} and $\gamma\approx\pi/6$ \cite{bon} in the rotational
energy operator. Here we use Davidson potential for $V(\beta_2)$ and
$V(\gamma)$ \cite{bonatsos1}.

In the free triaxiality approximation the rotational energy operator has the form
\begin{equation}
\hat{T}_{rot}=\frac{1}{4}\sum^{3}_{\lambda=1}\frac{\hat{I}^2_{\lambda}}{\left[sin\left(\gamma-\frac{2\pi}{3}\lambda\right)\right]^2},
\label{rotat}
\end{equation}
expands into a power series ($\gamma$-$\gamma_0$):
\begin{equation}
\hat{T}_{\rm rot}=\hat{T}_{\rm rot}(\gamma_0)+\frac{\partial\hat{T}_{\rm rot}}{\partial\gamma}|_{\gamma=\gamma_0}(\gamma-\gamma_0)+
\frac{1}{2}\frac{\partial^2\hat{T}_{\rm rot}}{\partial\gamma^2}|_{\gamma=\gamma_0}(\gamma-\gamma_0)^2+....
\label{serie}
\end{equation}
 where $\gamma_0$ - the parameter of transverse deformations of the nucleus surface is in the ground state.
 This approach, in contrast to the work \cite{iach,bon}
 allows you to take into account the full range of changes in $\gamma$ variable ($0\leq\gamma\leq\pi/3$).
\section{Wave functions}
Wave functions having the following form
$\Psi_{n_\gamma n_{\beta_2} IM\tau}(\beta_2,\gamma,\theta_i)$ \cite{max}:
\begin{equation}
\Psi_{n_\gamma n_{\beta_2} IM\tau}(\beta_2,\gamma,\theta_i)=F_{n_{\beta_2}}(\beta_2)\xi_{n_\gamma}(\gamma)\phi_{IM\tau}(\theta_i),
\label{psi}
\end{equation}
here
\begin{equation}
\xi_{n_\gamma}(\gamma)=\frac{N_{\gamma_{0}}\gamma^{p+1} }{\sqrt{|\sin 3\gamma|}}\exp\left(-\frac{\gamma^2}{2b^2_{\gamma_{0}}}\right)L^{p+1/2}_{n_\gamma}\left(\frac{\gamma^2}{b^2_{\gamma_{0}}}\right),
\label{xi}
\end{equation}
\begin{equation}
F_{n_{\beta_2}}(\beta_2)=N_{\beta_{0}}{\beta_2}^{q+1}\exp\left(-\frac{{\beta^2_2}}{2b^2_{\beta_{20}}}\right)L^{q+1/2}_{n_{\beta_2}}\left(\frac{\beta^2_2}
{b^2_{\beta_{20}}}\right),
\label{fnb}
\end{equation}
\begin{equation}
\phi_{IM\tau}(\theta_i)=\sqrt{\frac{2I+1}{16\pi^2(1+\delta_{0K})}}\left[D^M_{IK}(\theta_i)+(-1)^ID^I_{M,-K}(\theta_i)\right]A^\tau_{IK},
\end{equation}
where  $n_{\beta_2}$ and  $n_{\gamma}$  -- quantum numbers of
$\beta_2$- and $\gamma$- vibrations, respectively;
$\varepsilon_{I\tau}$ -- eigenvalues of the triaxial rotator; the
index $\tau$ enumerates the wave functions related to the same $IM$
\cite{davydov};
$\mu^4_{\beta_2}=\hbar^{2}/(B_2C_{\beta_2}\beta_{20}^{4})$ and
$\mu^4_{\gamma}=\hbar^{2}/(B_2C_{\gamma}\gamma_{0}^{4})$ --
$"$nonadiabacity$"$ dimensionless parameters with respect to
$\beta_2$- and $\gamma$-vibrations \cite{davyd}, respectively;
$C_{\gamma}$ and $C_{\beta_2}$ -- elasticity constants of the nucleus with respect to $\gamma$- and $\beta_2$- vibrations, respectively;
$N_{\gamma_{0}}$ and $N_{\beta_{20}}$ -- normalization coefficients;
$\gamma_0$ and  $\beta_{20}$ -- parameters  of $\gamma$-  and
$\beta_2$- vibrations in ground state, respectively; $K$
--projection of the spin onto the axial axis in the coordinate
system associated with the nucleus; $M$ -- spin projection onto the
axial axis in the laboratory coordinate system;
$p=\sqrt{\frac{1}{4}+\mu^{-4}_{\gamma_{0}}}-\frac{1}{2}$;
$q=-\frac{1}{2}+\sqrt{\varepsilon_{I\tau}+\mu^{-4}_{\beta_{2}}}$;
$L^{p+\frac{1}{2}}_{n_\gamma}(\gamma^2/b^2_{\gamma_{0}})$ and
$L^{q+1/2}_{n_{\beta_2}}({\beta^2_2}/b^2_{\beta_{20}})$ -- Laguerre
polynomials; $b_{\gamma_{0}}=\gamma_0\mu_{\gamma_{0}}$;
$b_{\beta_{20}}=\beta_{20}\mu_{\beta_2}$;
 $D^I_{IK}(\theta_i)$ -- Wigner function; $\theta_i$ -- Euler angles; $\delta_{0K}$ -- Kronecker symbol.

\section{Energy levels of collective states}
In the case of zero approximation in the expansion of the rotational
energy operator  in the framework of the free triaxiality model
\cite{davyd}, we obtained the following expressions for the energy
spectrum \cite{max}
\begin{equation}
E_{0n_\gamma n_{\beta_2} I\tau}=2n_{\beta_2}+\sqrt{4n_\gamma\frac{\mu^{-2}_{\gamma}}{\gamma^2_0}+\varepsilon_{I\tau}+\mu^{-4}_{\beta_2}+\frac{1}{4}},
\label{nonax}
\end{equation}

\section{The corrections to the energy of the excited levels}
Consider the corrections to the energy of the excited levels of the ground-, $\beta$-, and $\gamma$-bands of the second $\varepsilon_1$ and third terms $\varepsilon_2$ in the expansion of the rotational energy operator (\ref{serie}) \cite{max,por }:
\begin{equation}
\varepsilon_1=N^2_{\gamma_{0}}<\gamma^{p+1}exp(-\frac{\gamma^2}{2b^2_{\gamma_0}})\phi_{IM\tau}(\theta_i)|\frac{\partial\hat{T}_{\rm rot}}{\partial\gamma}|_{\gamma=\gamma_0}(\gamma-\gamma_0)\phi_{IM\tau}(\theta_i)\gamma^{p+1}exp(-\frac{\gamma^2}{2b^2_{\gamma_0}})>=
\label{deriv1}
\end{equation}
$$
=<\phi_{IM\tau}(\theta_i)|\frac{\partial\hat{T}_{\rm rot}}{\partial\gamma}|_{\gamma=\gamma_0}\phi_{IM\tau}(\theta_i)>N^2_{\gamma_{0}}\left[\int_0^{\frac{\pi}{3}}\gamma^{2p+2} exp\left(-\frac{\gamma^2}{b^2_{\gamma_0}}\right)(\gamma-\gamma_0)d\gamma\right]^2,
$$

$$
<\phi_{IM\tau}(\theta_i)|\frac{\partial\hat{T}_{\rm rot}}{\partial\gamma}|\phi_{IM\tau}(\theta_i)>=
$$
$$
=-\sum_{K\geq0}A^\tau_{IK}A^{\tau'}_{IK}\left\{\left[\frac{2\cos(\gamma+2\pi/3)}{\sin^3(\gamma+2\pi/3)}+\frac{2\cos(\gamma-2\pi/3)}{\sin^3(\gamma-2\pi/3)}\right]\times \right.
$$
$$
\times\left.
\frac{I(I+1)-K^2}{8}+\frac{2\cos\gamma}{4\sin^3\gamma}K^2\right\}+
$$
$$
+\frac{1}{16}\left[\frac{2\cos(\gamma+2\pi/3)}{\sin^3(\gamma+2\pi/3)}-\frac{2\cos(\gamma-2\pi/3)}{\sin^3(\gamma-2\pi/3)}\right]\times
$$
$$
\times\sum_{K\geq0}\left(A^\tau_{IK+2}A^{\tau'}_{IK}+
A^\tau_{IK}A^{\tau'}_{IK+2}\right)\frac{[1+(-1)^I\delta_{K0}]f(K)}{\sqrt{1+\delta_{K0}}},
$$

\begin{equation}
\varepsilon_2=N^2_{\gamma_{0}}<\gamma^{p+1} exp(-\frac{\gamma^2}{2b^2_{\gamma_0}})\phi_{IM\tau}(\theta_i)|\frac{\partial^2\hat{T}_{\rm rot}}{2\partial\gamma^2}|
_{\gamma=\gamma_0}(\gamma-\gamma_0)^2\phi_{IM\tau}(\theta_i)\gamma^{p+1} exp(-\frac{\gamma^2}
{2b^2_{\gamma_0}})>=
\label{deriv2}
\end{equation}
$$
=<\phi_{IM\tau}(\theta_i)|\frac{\partial^2\hat{T}_{\rm rot}}
{\partial\gamma^2}|_{\gamma=\gamma_0} \phi_{IM\tau}
(\theta_i)>N^2_{\gamma_{0}}\left[\int_0^{\frac{\pi}{3}}\gamma^{2p+2} exp\left(-\frac{\gamma^2}{b^2_{\gamma_0}}\right)(\gamma-\gamma_0)^2d\gamma\right]^2,
$$

$$
<\phi_{IM\tau}(\theta_i)|\frac{\partial^2\hat{T}_{\rm rot}}{\partial\gamma^2}|\phi_{IM\tau}(\theta_i)>=
$$
$$
=\sum_{K\geq0}A^\tau_{IK}A^{\tau'}_{IK}
\left\{\left[2a(3a-2)+2b(3b-2)\right]\times \right.
$$
$$
\times\left.
\frac{I(I+1)-K^2}{8}+\frac{2c(3c-2)K^2}{4}\right\}
+\frac{\left[2a(3a-2)-2b(3b-2)\right]}{16}\times
$$
$$
\times\sum_{K\geq0}\left(A^\tau_{IK+2}A^{\tau'}_{IK}+
A^\tau_{IK}A^{\tau'}_{IK+2}\right)\frac{[1+(-1)^I\delta_{K0}]f(K)}
{\sqrt{1+\delta_{K0}}},
$$
where
\begin{eqnarray}
a=\frac{1}{\sin^2(\gamma-2\pi/3)},
b=\frac{1}{\sin^2(\gamma+2\pi/3)},
c=\frac{1}{\sin^2\gamma},\\
f(K)=\sqrt{(I+K+2)(I+K+1)(I-K-1)(I-K)}.
\end{eqnarray}

Taking into account (\ref{nonax}), (\ref{deriv1}), (\ref{deriv2}), the expression for the energy spectrum takes the following form:
\begin{equation}
E_{n_\gamma n_{\beta_2} I\tau}=\hbar\omega
\left(E_{0n_\gamma n_{\beta_2} I\tau}+\varepsilon_1+\varepsilon_2\right).
\label{nonax1}
\end{equation}
where $\hbar\omega$ is the energy multiplier.

Note that the expression (\ref{nonax1}) depends on the parameters
 $\hbar\omega$, $\gamma_{0}$, $\mu_{\beta_2}$, $\mu_{\gamma}$.

\section{Result}
In the Table \ref{table1} the comparison of theoretical and
experimental values  \cite{nndc} is presented  for nuclei lanthanide and actinides of ratio R$_{4/2}$=E($4_1$)/E($2_1$),
R$_{0_\beta}$=E(0$_\beta$)/ E($2_1$) and
R$_{2_\gamma}$=E($2_\gamma$)/E($2_1$) respectively. As well as, the
values of the parameters $\hbar\omega$, $\mu_{\beta_2}$, $\mu_\gamma$,
$\gamma_0$ and  RMS (in keV). Note that the triaxial parameter
$\gamma_0$ takes a value 9$^0\div$14.4$^0$ for lanthanide,
7.8$^0\div$10.6$^0$ for actinides. The quadrupole deformation
parameter $\mu_{\beta_2}$=0.22$\div$0.39.

 Note at 2.7$<R_{4/2}<$10/3, the collective behavior of the energy spectrum of the levels will be rotational. And at 2$<R_{4/2}<$2.4 it will be vibrational \cite{minkov11}.  In our case R$_{4/2}$=2.93$\div$3.31, i.e. the energy spectrum of the levels is rotational.
 
\begin{figure}
	\includegraphics[height=8cm, width=13cm]{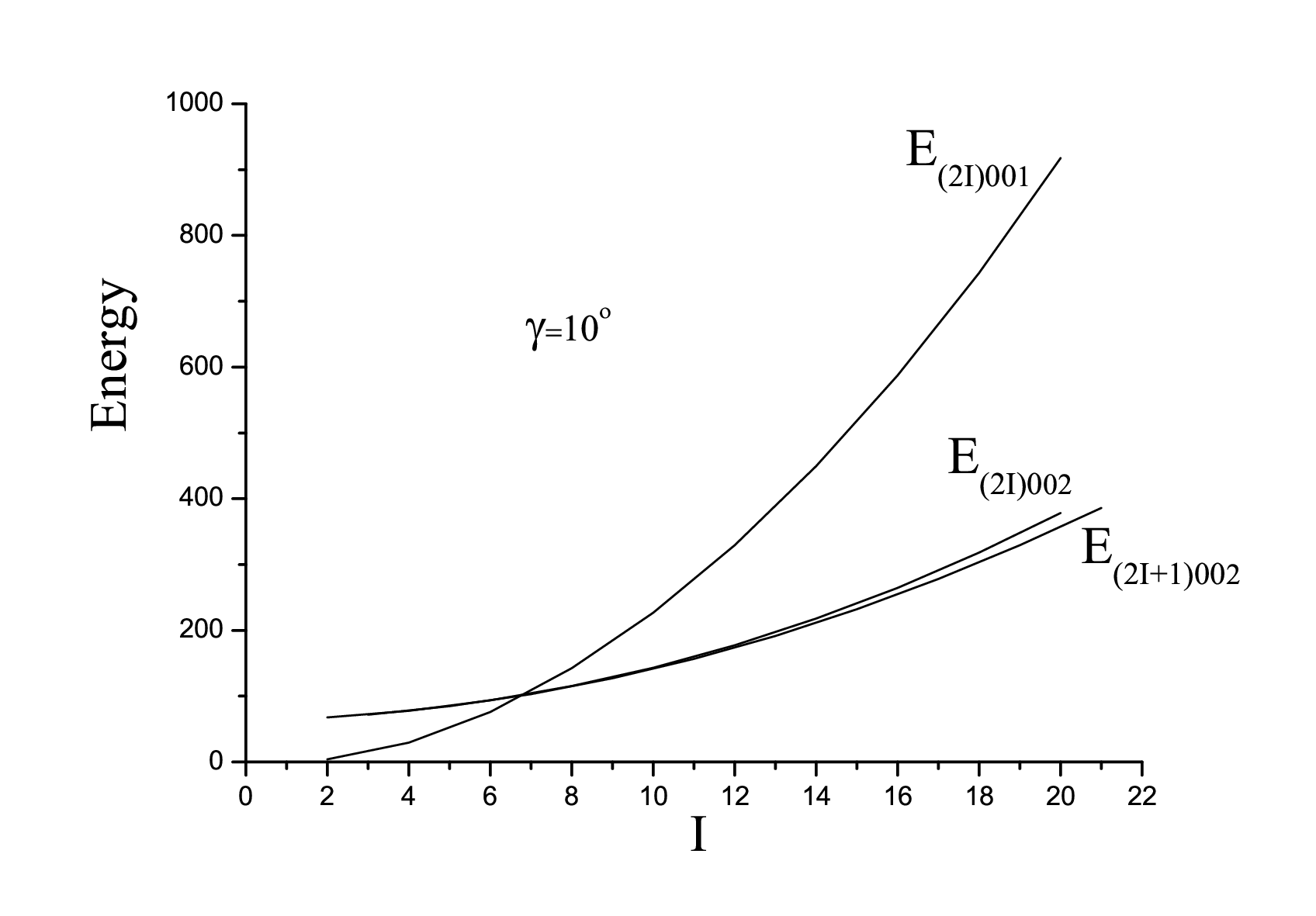}
	\caption{Behavior energy levels (\ref{nonax1}) gsb [(2I)001] and $\gamma$-band [(2I)002 for even spins, (2I+1)002 for odd spins] for values of the parameters  $\mu_{\gamma_0}$=2 and $\mu_{\beta_2}$=0.3. Here and in all figures I=1,2,3...}
	\label{energy}
\end{figure}

In fig.\ref{energy} the behavior energy levels of 
(\ref{nonax1}) ground state band (gsb) [(2I)001] and $\gamma$-band is presented
[(2I)002 for even spins, (2I+1)002 for odd spins] for values of the parameters $\gamma=10^0$, $\mu_\gamma$=2 and $\mu_{\beta_2}$=0.3
\footnote{Furthermore, we will study the reduced \emph{E}2-transitions
probabilities, where there are \emph{E}2-transitions between these states, so we have presented these states separately.}. Note that here we observe bandcrossing at values I=7. This means that at large spins the behavior of the moment of inertia of nucleus will be anomalous \cite{mottel,bonatsos}. Experimental data provide convincing evidence that reverse bending is a manifestation of the intersection of the ground state band with another rotational band that has a large moment of inertia. $"$Coriolis antipairing$"$ and $"$Coriolis decoupling$"$ have emerged as two mechanisms that most likely produce a band that crosses the ground state band \cite{simon}.

In the Table \ref{table220}  shows the sequence of  the energy levels of the ground and $\gamma$-bands for different values of the triaxiality parameter $\gamma_0$. It can be seen from this table that for small values of the triaxiality parameter of the energies of the levels of the ground and $\gamma$ bands are very different. In this case these bands have no interactions. As the values of the triaxiality parameter $\gamma_0$ increase, the energies of the levels of the ground and $\gamma$-bands take on close values at I=12 and $\gamma_0$=5$^0$. And for large values of the triaxiality parameter $\gamma_0$, the energies of the levels of the ground and $\gamma$-bands take close values, already at I=4 and $\gamma_0$=30$^0$. Since with increasing energy excitation or impulse momentum, the value of the moment of inertia of the nucleus increases. This is due to the decrease in the pairing forces with increasing energy. This happens when energy levels with the same spins from different ones are close or there is an intersection of the bands. Thus the increase in the value of the moment of inertia of the nucleus is the result of the manifestation of interaction or crossing of the bands.

In the Table \ref{table2} the experimental values\cite{nndc} excited states for the $\gamma$-band of even-even nuclei are presented. Following some even-even nuclei has  high spin states in $\gamma$-band: $^{156}$Dy
($I_{max}$=13 and $\gamma$=13.9$^0$), $^{156}$Gd ($I_{max}$=13 and
$\gamma$=10.4$^0$), $^{164}$Er ($I_{max}$=13 and $\gamma$=13$^0$),
$^{232}$Th ($I_{max}$=12 and $\gamma$=9.2$^0$) and $^{238}$U
($I_{max}$=26 and $\gamma$=7.9$^0$). Simultaneously, the explanation of
the excited states of the ground-state-band and $\gamma$-band of the above-mentioned nuclei at high spins (I$>$7) is inappropriate.

\section{Reduced \emph{E}2-transitions  probabilities between excited collective states of triaxial even-even nuclei}
The reduced \emph{E}2-transitions  probabilities between the
$n'_\gamma n'_{\beta_2} I'\tau'$ and $n_\gamma n_{\beta_2} I \tau$
\cite{max} states are given by
\begin{equation}
B(\emph{E}2;n_\gamma n_{\beta_2} I\tau\rightarrow n'_\gamma n'_{\beta_2} I'\tau')=\frac{5}{16\pi(2I+1)}\sum_{MM'\mu}|<n'_\gamma n'_{\beta_2} I'\tau'|Q_{2\mu}|n_\gamma n_{\beta_2} I\tau>|^2,
\end{equation}
where
\begin{equation}
Q_{2\mu}=\frac{\beta_2}{\beta_{20}}\left[Q_0D^2_{\mu0}\cos\gamma+Q_2\left(D^2_{\mu2}+D^2_{\mu,-2}\right)\frac{\sin\gamma}{\sqrt{2}}\right],
\end{equation}
where $Q_0$ is the quadrupole moment about the symmetry axis, $Q_2$
is the measure of the shape asymmetry about the symmetry axis
\cite{bohr}. For a system with an ellipsoidal shape, the quadrupole
moment of the nucleus can be expressed in terms of two internal
quadrupole moments $Q_0$ and $Q_2$. By the by \cite{bohr}
$$Q_2=\frac{\tan\gamma_0}{\sqrt{2}}Q_0.$$
Reduced \emph{E}2-transitions  probabilities between $i\equiv\{n_\gamma n_{\beta_2} I\tau\}$ and $f\equiv\{n'_\gamma n'_{\beta_2} I'\tau'\}$ can be represented as
\begin{equation}
B(\emph{E}2;i\rightarrow f)=B_a(\emph{E}2;I\tau\rightarrow I'\tau')S_{if\beta_2}^{2}S_{if\gamma}^{2}, \label{31}
\end{equation}
where $B_a(\emph{E}2;I\tau\rightarrow I'\tau')$ -- \emph{E}2-transitions  probabilities in excited states of a symmetric rotator
\cite{davyd} and
\begin{equation}
S_{if\beta_2}=\int_{0}^{\infty}F_{i}(\beta_2)\frac{\beta_2}{\beta_{20}}F_{f}(\beta_2)\beta^4_2d\beta_2,
\label{32}
\end{equation}
which takes into account the deformability of an even-even nucleus. The matrix element $S_{if\gamma}$ has the form
$$
S_{if\gamma}=\int_{o}^{\infty}\xi_{i}(\gamma)\phi_i(\theta_i)\left\{Q_0D^2_{\mu0}\cos\gamma+
Q_2\left(D^2_{\mu2}+D^2_{\mu,-2}\right)\frac{\sin\gamma}{\sqrt{2}}\right\}\times
$$
\begin{equation}
\times\xi_{f}(\gamma)\phi_f(\theta_i)|\sin 3\gamma|\sin\theta_2d\gamma d\theta_1d\theta_2d\theta_3.
\label{321}
\end{equation}
takes into account the influence of $\gamma$-oscillations on the
reduced \emph{E}2-transitions  probabilities. Given the expression
(\ref{xi}) and the reduced matrix elements defined by formula (10.8)
in \cite{davyd}. The expression (\ref{321}) can be rewritten as
follows:
$$
S_{if\gamma}=\left\{Q_0\sqrt{2I+1}(I2K0|I'K')
J_1+Q_2\sqrt{\frac{2I+1}{2}}\left[\sqrt{\frac{1+\delta_{K0}}{1+\delta_{K'0}}}(I2K2|I'K')+\right.\right.
$$
\begin{equation}
\left.\left.+\sqrt{\frac{1+\delta_{K'0}}{1+\delta_{K0}}}(I2K-2|I'K')\right]J_2\right\}
\frac{2}{\lambda[\pi^2/(9b^2_{\gamma_{0}}),p+3/2]b_{\gamma_{0}}^{p+p'+3}},
\end{equation}
where $\lambda(x,a)$ is an incomplete Gamma function; $(I2K0|I'K')$--Clebsch-Gordon coefficients.
\begin{equation}
J_1=\int_{0}^{\pi/3}\gamma^{p+p'+2}
\exp\left(-\frac{\gamma^2}{b^2_{\gamma_{0}}}\right)\cos\gamma d\gamma,
\end{equation}

\begin{equation}
J_2=\int_{0}^{\pi/3}\gamma^{p+p'+2}
\exp\left(-\frac{\gamma^2}{b^2_{\gamma_{0}}}\right)\sin\gamma d\gamma.
\end{equation}
The integrals $J_1$ and $J_2$ are calculated numerically.

The general expression  for matrix elements $S_{if\beta_2}$ is very
cumbersome. Therefore, we consider special cases. For the reduced \emph{E}2-transitions  probabilities in the ground-band, which
correspond to the values: $n_{\beta_2}$=0, $n^{'}_{\beta_2}$=0,
$n_\gamma$=0 and $n^{'}_\gamma$=0, expression (\ref{32}) becomes:
\begin{equation}
S_{if\beta_2}=\mu_{\beta_2}\sqrt{\frac{1}{{\Gamma(q+\frac{7}{2})}{\Gamma(q'+\frac{7}{2})}}}\Gamma\left(\frac{q+q'+7}{2}\right).
\label{ground}
\end{equation}
where $\Gamma(x)$--Gamma function.

Matrix elements (\ref{32}) of the reduced \emph{E}2-transitions  probabilities between the ground and $\beta$-bands, corresponding to the
values $n_{\beta_2}$=1, $n'_{\beta_2}$=0, $n_\gamma$= 0 and
$n'_\gamma$=0 have the form:
$$
S_{if\beta_2}=\frac{\mu_{\beta_2}}{4}\sqrt{\frac{1}{(q+\frac{15}{2})(q'+\frac{15}{2})\Gamma(q+\frac{7}{2})\Gamma(q'+\frac{7}{2})}}\Gamma\left(\frac{q+q'+8}{2}\right)\times
$$
\begin{equation}
\times [(q+q'+10)(q+q'+8)-2(q+q'+3)(q+q'+8)+(2q+3)(2q'+3)].
\label{beta}
\end{equation}

Matrix elements (\ref{32}) of the reduced \emph{E}2-transitions  probabilities in the $\beta$-band, corresponding to the values $n_{\beta_2}$=1, $n'_{\beta_2}$=1, $n_\gamma$= 0 and $n'_\gamma$=0 have the form:
\begin{equation}
S_{if\beta_2}=\frac{\mu_{\beta_2}}{2}\sqrt{\frac{1}{(q+\frac{15}{2})\Gamma(q+\frac{7}
{2})\Gamma(q'+\frac{7}{2})}}\Gamma\left(\frac{q+q'+8}{2}\right)(q-q'-5).
\label{bg}
\end{equation}
The intra/inter-band reduced \emph{E}2-transition probabilities are
expressed in terms of four parameters: $\mu_{\beta_2}$, $\mu_\gamma$,
$\gamma_0$  and $Q_0$. Note that the parameters $\mu_{\beta_2}$,
$\mu_\gamma$ and $\gamma_0$ are obtained from the description of the
energy levels of excited states. The  parameter $Q_0$ were obtained
from a comparison of the calculated intra-/inter-band reduced
\emph{E}2-transitions  probabilities with experimental data.

It is convenient to consider the ratio of the reduced \emph{E}2-transitions  probabilities  to the reduced \emph{E}2-transition  probability from the state of the first excited level
with spin $2^{+}$ to the ground state, which is given by:

\begin{equation}
b(E2;n'_\gamma n'_{\beta_2} I_i\tau'\rightarrow n_\gamma n_{\beta_2} I_f\tau)=\frac{B(E2;n'_\gamma n'_{\beta_2} I_i\tau'\rightarrow
n_\gamma n_{\beta_2} I_f\tau)}{B(E2;0021\rightarrow 0001)}.
\label{ratio22}
\end{equation}

The basic shape properties of atomic nuclei are described through
parameter of quadrupole deformation $\beta_2$ and axial-asymmetry
parameter $\gamma_0$ \cite{bohr}. The parameter $\beta_2$ is related
to the intrinsic quadrupole moment and describes the magnitude of
departure from the spherical shape along the principal axis in the
intrinsic frame. 

Here we will study various permitted \emph{E}2-transitions, depending on
the triaxiality parameter $\gamma_0$. At the same time, we take into
account the entire range of changes in the parameter $\gamma_0$. All
these ratios of the reduced \emph{E}2-transitions probabilities
(\ref{ratio22}) are  presented in figs.\ref{fig1}-- \ref{fig12}. The
behavior of these ratios of the reduced \emph{E}2-transitions probabilities
in more large values of the parameter $\gamma_0$ to explain is
difficult. The major reason of this behavior is an anomalous
behavior moment of inertia of the nucleus at large values of the
parameter $\gamma_0$.
\subsection{Intra-band reduced \emph{E}2-transition probabilities}
\begin{figure}
\begin{center}
\includegraphics[height=8cm, width=12cm]{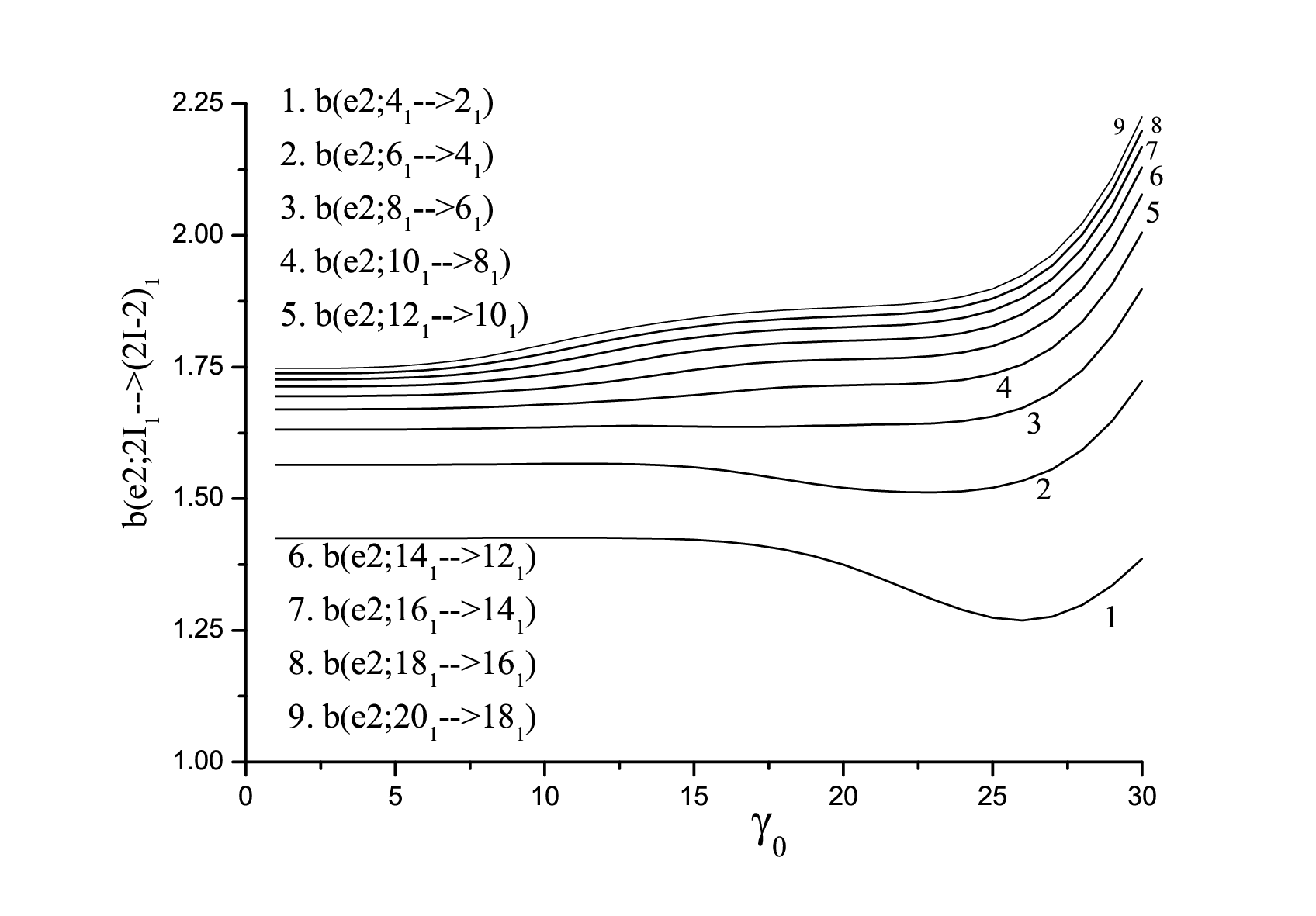}
\caption{Dependency reduced b(E2,$2I_1\to (2I-2)_1$)-transition probabilities on the parameter $\gamma$ at $Q_0$=9, $Q_2$=2, $\mu_{\gamma_0}$=2, $\mu_{\beta_2}$= 0.3. Here and in all figures I=1,2,3...}
\label{fig1}
\end{center}
\end{figure}

\begin{figure}
\begin{center}
\includegraphics[height=8cm, width=12cm]{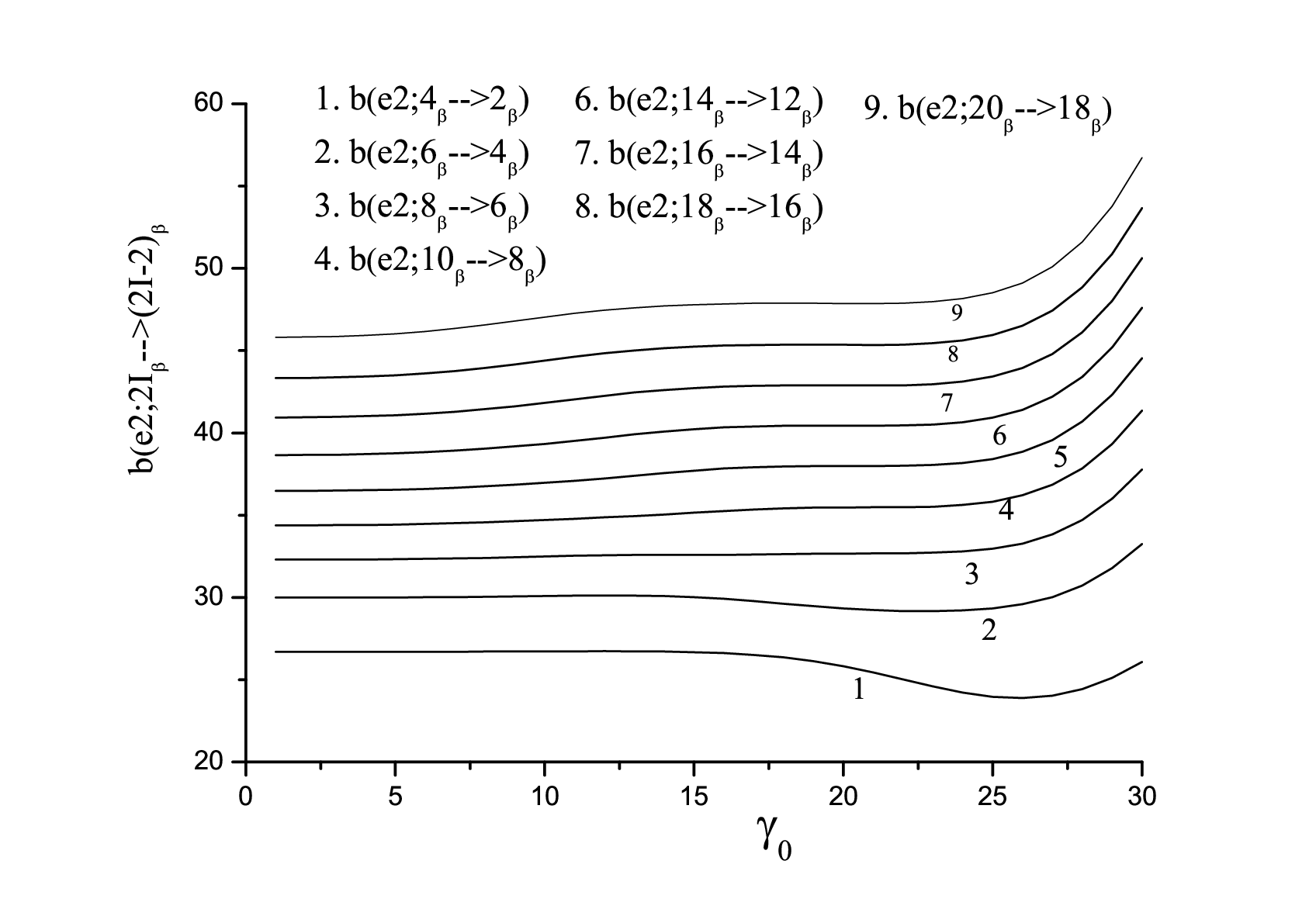}
\caption{The same as in Fig. \ref{fig1}, but for b(E2,$2I_\beta\to (2I-2)_\beta$).}
\label{fig2}
\end{center}
\end{figure}

\begin{figure}
\begin{center}
\includegraphics[height=8cm, width=12cm]{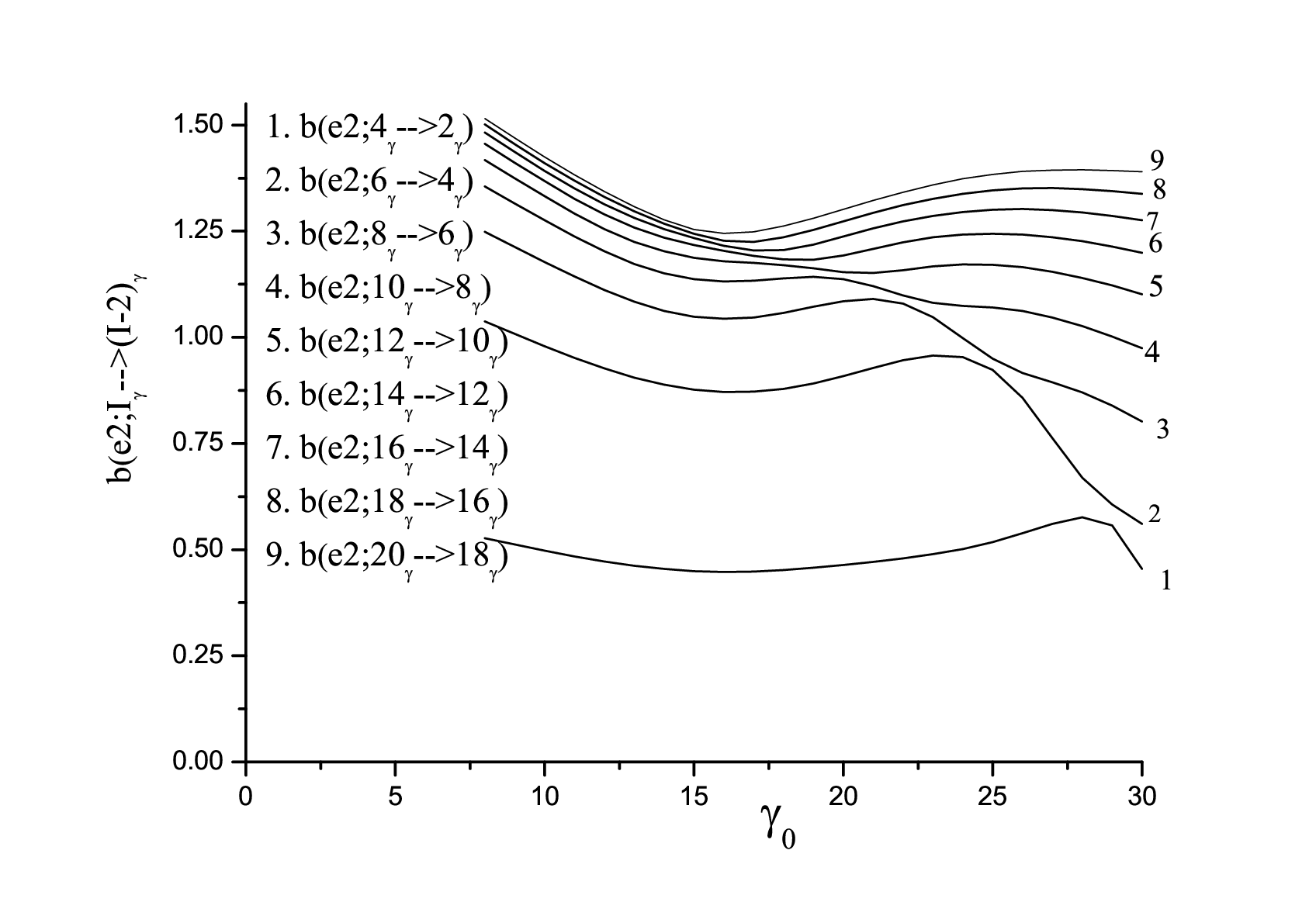}
\caption{The same as in Fig. \ref{fig1}, but for  b(E2,$2I_\gamma\to (2I-2)_\gamma$) between even spins.}
\label{fig3}
\end{center}
\end{figure}

\begin{figure}
\begin{center}
\includegraphics[height=8cm, width=12cm]{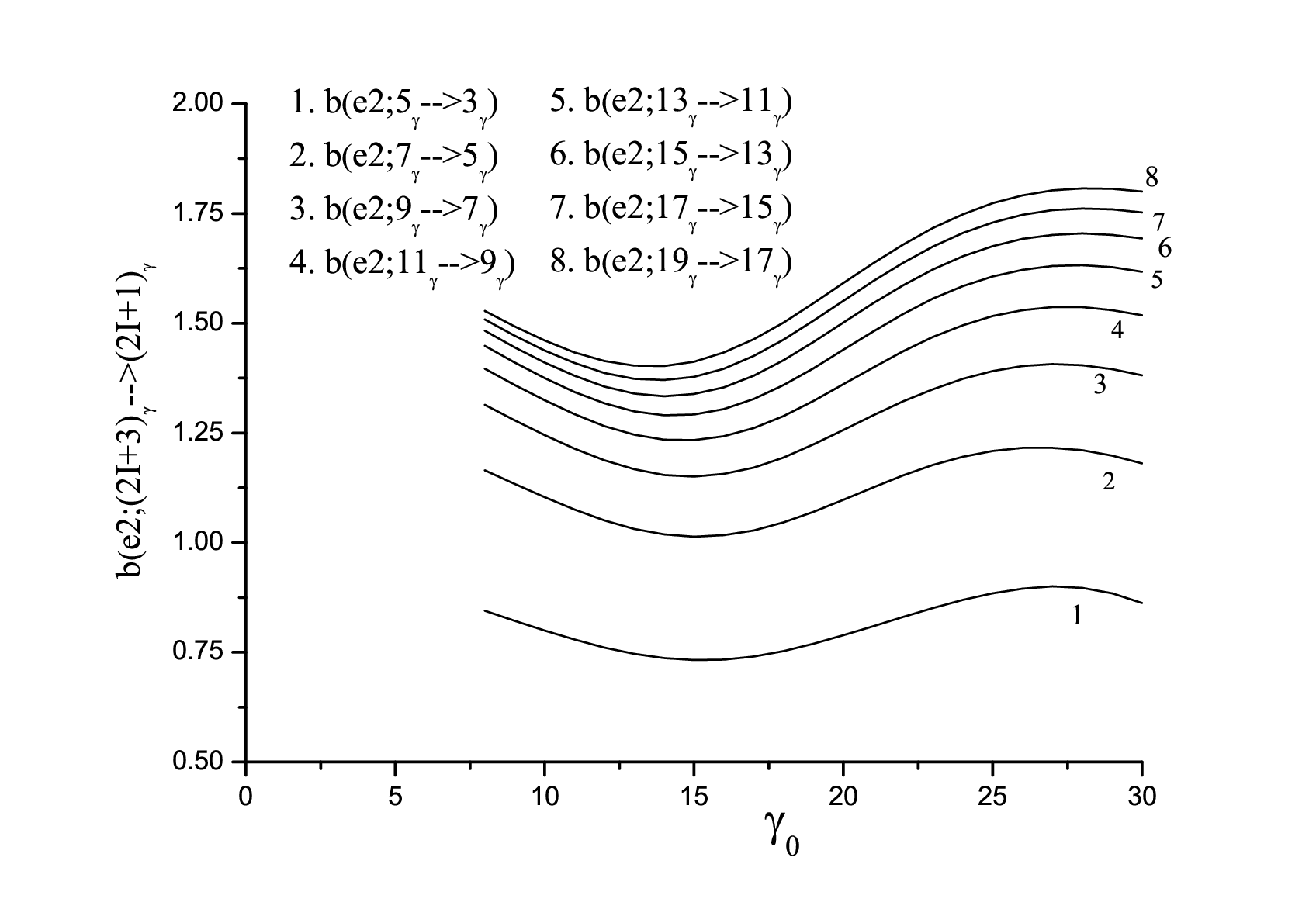}
\caption{The same as in Fig. \ref{fig1}, but for b(E2,$(2\cdot I+3)_\gamma\to (2\cdot I+1)_\gamma$) between odd spins.}
\label{fig4}
\end{center}
\end{figure}

\begin{figure}
\begin{center}
\includegraphics[height=8cm, width=12cm]{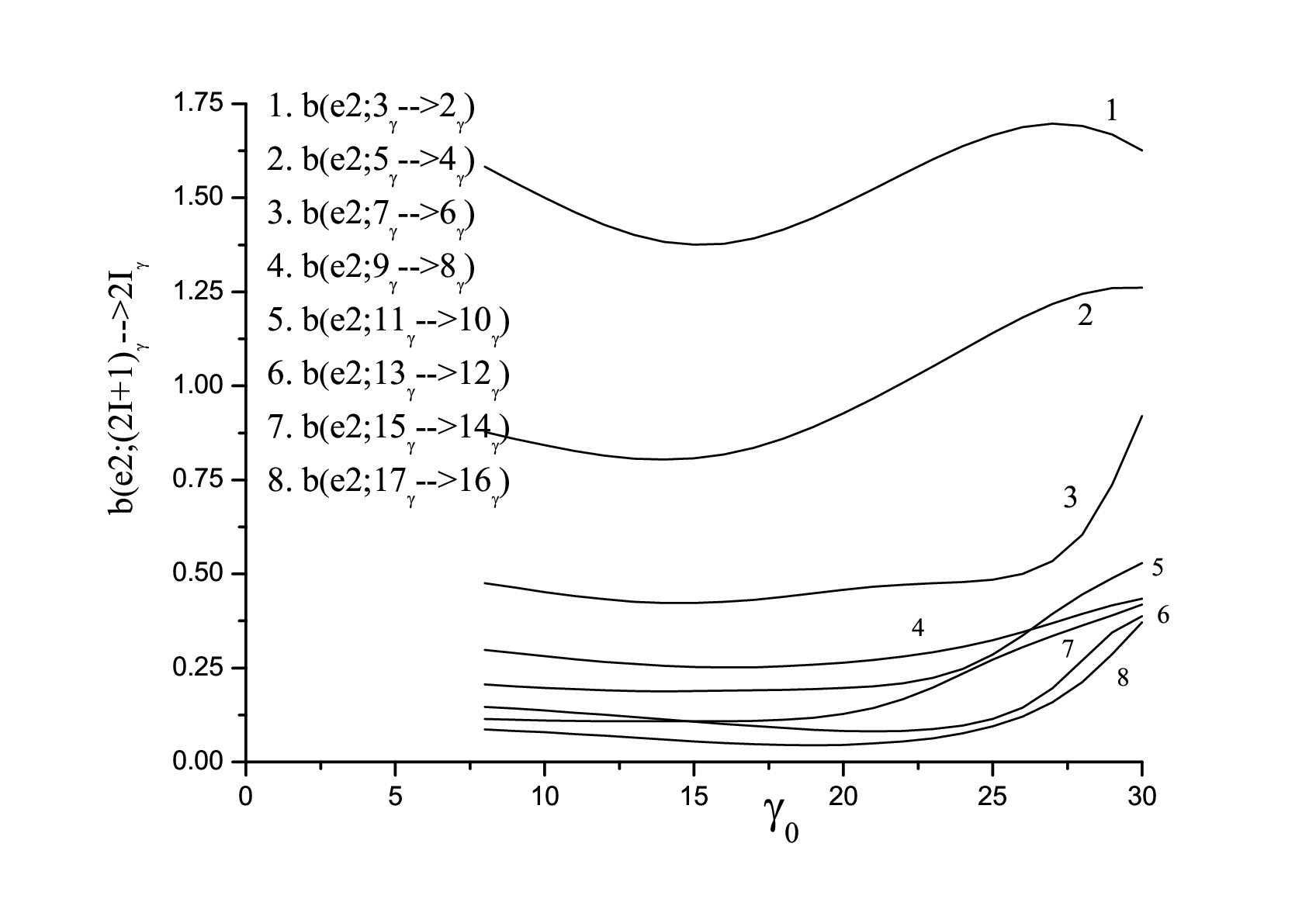}
\caption{The same as in Fig. \ref{fig1}, but for b(E2,$( 2I+1)_\gamma\to 2I_\gamma$).}
\label{fig5}
\end{center}
\end{figure}

In fig. \ref{fig1}-\ref{fig5} the intra-band reduced \emph{E}2-transition probabilities ground-state-band, $\beta$- and $\gamma$-bands are presented. At the same time, in fig. \ref{fig3}-\ref{fig5} the \emph{E}2-transitions between even, odd and mixed spins  for $\gamma$-band separately are represented.

In fig. \ref{fig1} intra-band reduced \emph{E}2-transition probabilities in ground-state-band are presented. These transitions at values $0^0\leq\gamma_0\leq 20^0$ of the  parameter $\gamma_0$ are insensitive to changes in this parameter, while at values $20^0\leq\gamma_0\leq 30^0$ of the  parameter the sensitivity of the transitions increases. These transitions  in the collective excitation spectra of almost all heavy deformed even-even nuclei are observed, including high spin states. From fig. \ref{fig1} obviously, the influence deformation parameter  $\mu_{\beta_2}$ to \emph{E}2-transitions is not sensitive \footnote{Due to the volume of the article, these results are not presented here.}, because this band is mostly rotational.

Behavior intra-band reduced \emph{E}2-transition probabilities for
ground-state-band, $\beta$-band are similar, but values of the \emph{E}2-transition probabilities in this band are much larger than in the ground-state-band. These transitions at small values of the $\gamma_0$ parameter are also insensitive to changes of this parameter. And their amplitude values are slightly different on the variable parameter $\gamma_0$.  In fig. \ref{fig2} obviously, the influence $\beta_2$-deformation to \emph{E}2-transitions is more sensitive. The values of these \emph{E}2-transitions decreases with increasing deformation parameter $\mu_{\beta_2}$. Their amplitude values decreases with increasing deformation parameter $\mu_{\beta_2}$ $^\texttt{b}$.  b(E2,$2I_\beta\to (2I-2)_\beta$) transitions are observed in the collective excitation spectra of most heavy deformed
even-even nuclei, so this band is mostly vibrational-rotational.

Intra-band reduced \emph{E}2-transition probabilities in $\gamma$-band
between even spins have more complex behavior. These transitions at small values of the $\gamma_0$ parameter are mostly sensitive to changes of this parameter. In fig. \ref{fig3}
obviously, that the influence $\beta_2$-deformation to \emph{E}2-transitions
is more sensitive. The values of these \emph{E}2-transitions increases with increasing deformation parameter $\mu_{\beta_2}$  $^\texttt{b}$.  These transitions are observed in the collective excitation spectra of most heavy deformed even-even nuclei: $^{164,166,168}$Er \cite{nndc}, because this band is vibrational-rotational.

Intra-band reduced \emph{E}2-transition probabilities in $\gamma$-band
between odd spins have more complex behavior. In fig. \ref{fig4}
obviously, the influence $\beta_2$-deformation to \emph{E}2-transitions
is more sensitive. The values of these \emph{E}2-transitions increases with increasing deformation parameter $\mu_{\beta_2}$ $^\texttt{b}$.  These transitions are
observed in the collective excitation spectra of most heavy deformed
even-even nuclei: $^{166,168}$Er \cite{nndc}, because this band is vibrational-rotational.
 Note, from fig. \ref{fig3} and \ref{fig4} obviously, the behavior reduced \emph{E}2-transitions probabilities intra odd and even spins of the $\gamma$ band are
different.

On fig. \ref{fig5} the behavior of the intra-band reduced
\emph{E}2-transitions probabilities  between $(2I+1)_\gamma\to 2I_\gamma$ is
presented. B(E2,$( 2I+1)_\gamma\to 2I_\gamma$) are more  sensitive to changes of the parameter $\gamma_0$. These transitions are observed in the collective
excitation spectra of most heavy deformed even-even nuclei:
$^{152}$Sm and $^{166,168}$Er \cite{nndc}.

\subsection{Inter-band reduced \emph{E}2-transition probabilities}
\begin{figure}
\begin{center}
\includegraphics[height=8cm, width=12cm]{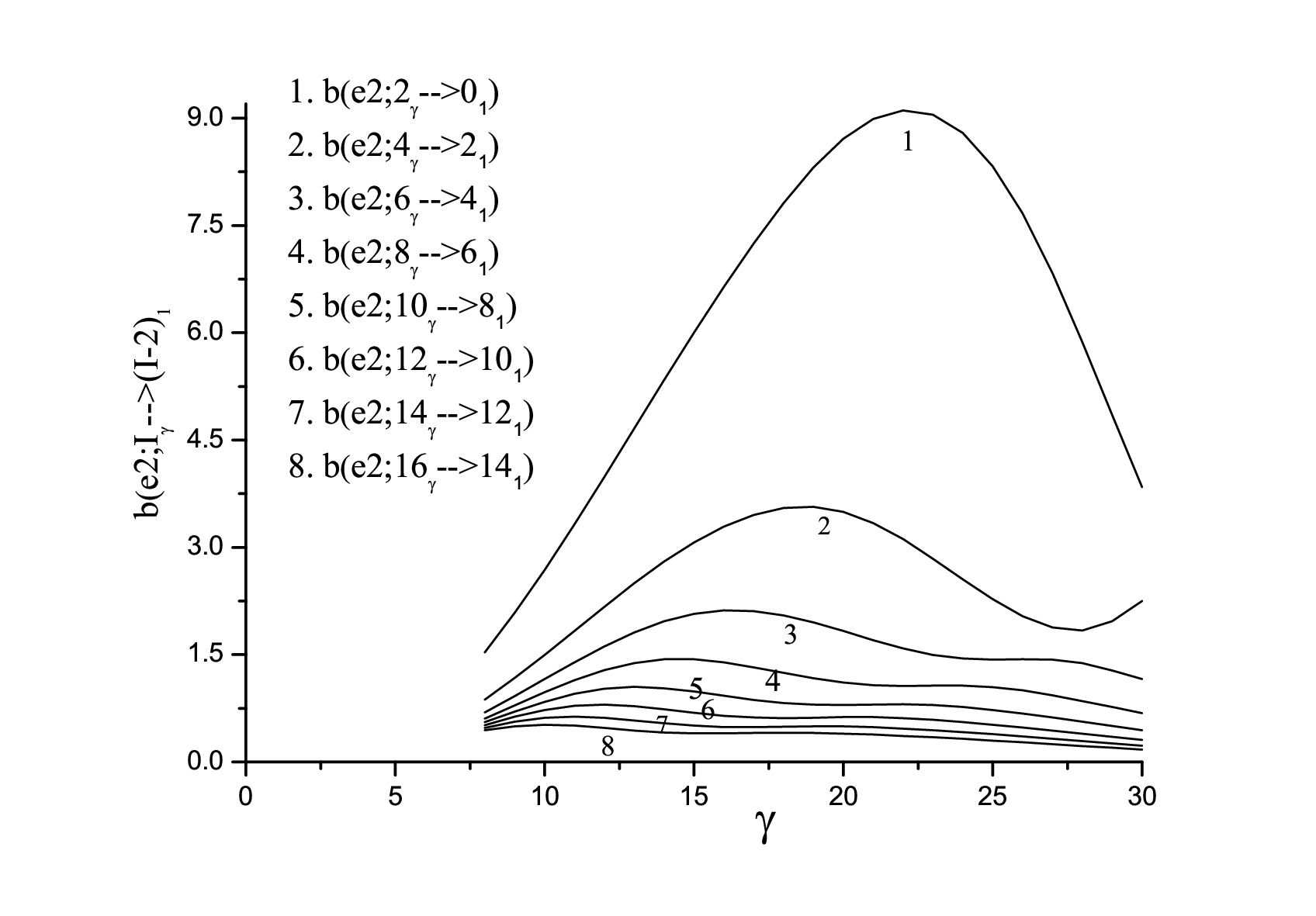}
\caption{The same as in Fig. \ref{fig1}, but for b(E2,$2I_\gamma\to (2I-2)_1$).}
\label{fig6}
\end{center}
\end{figure}

\begin{figure}
\begin{center}
\includegraphics[height=8cm, width=12cm]{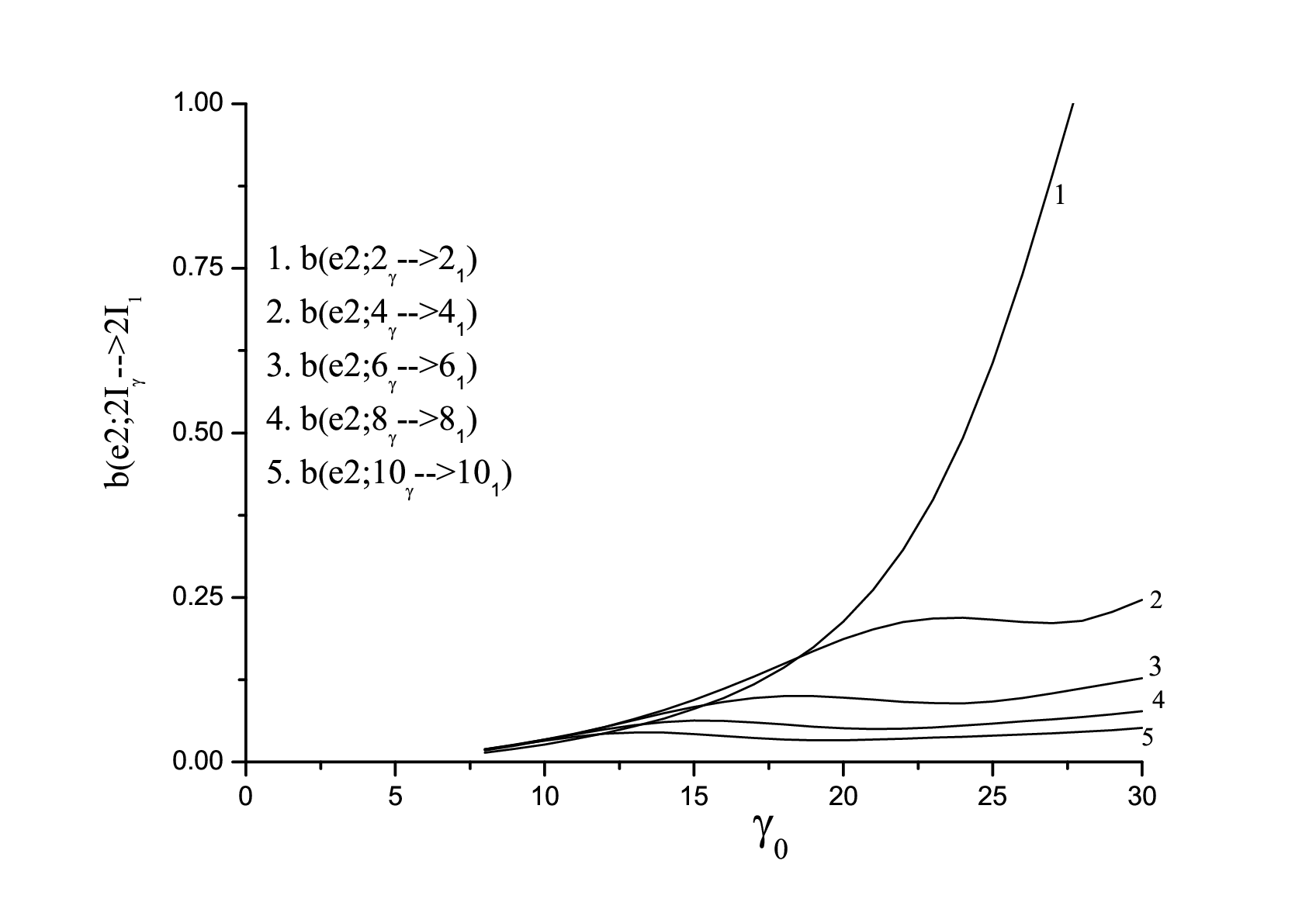}
\caption{The same as in Fig. \ref{fig1}, but for b(E2,$2I_\gamma\to 2I_1$).}
\label{fig7}
\end{center}
\end{figure}

\begin{figure}
\begin{center}
\includegraphics[height=8cm, width=12cm]{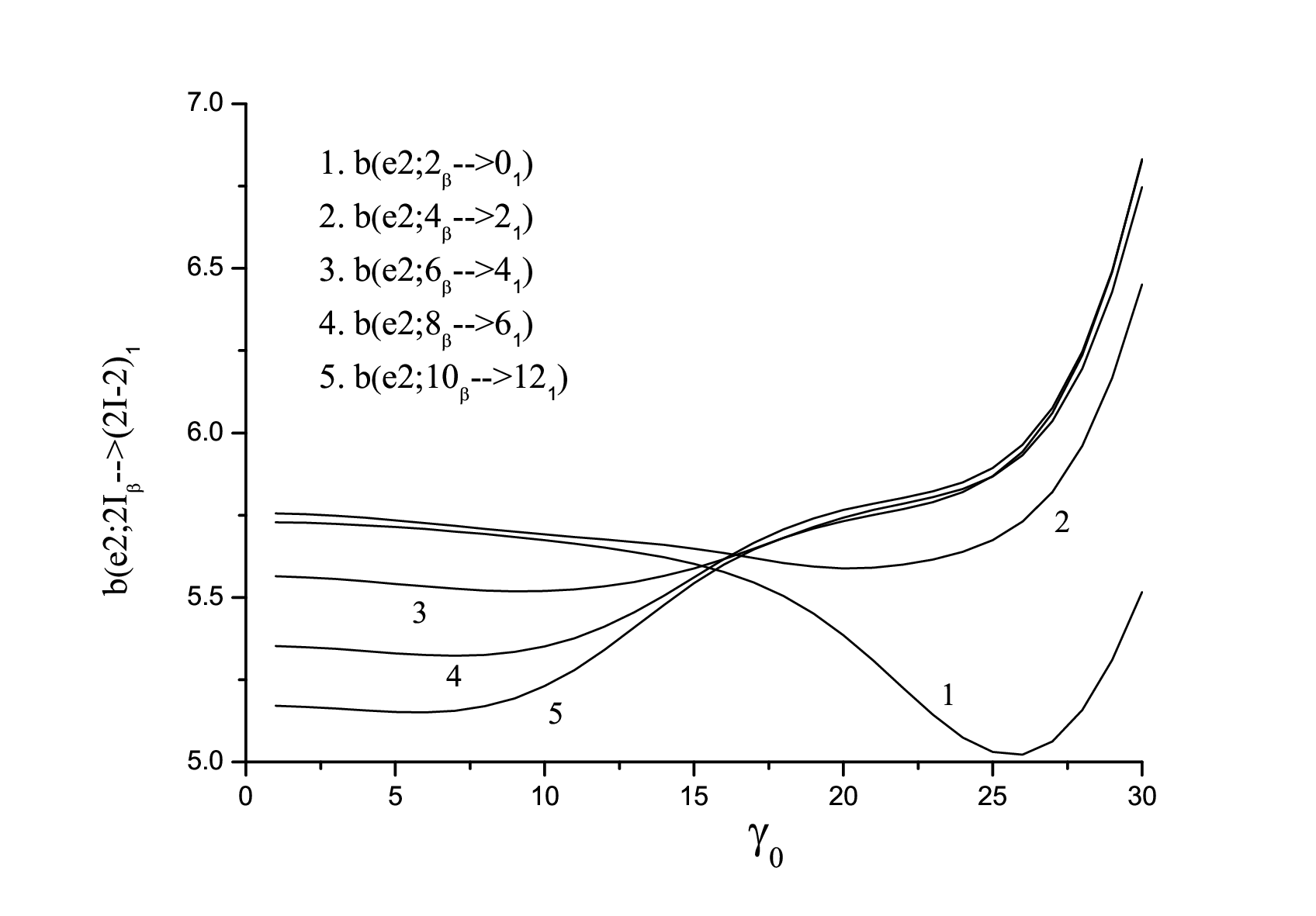}
\caption{The same as in Fig. \ref{fig1}, but for b(E2,$2I_\beta\to (2I-2)_1$).}
\label{fig8}
\end{center}
\end{figure}

\begin{figure}
\begin{center}
\includegraphics[height=8cm, width=12cm]{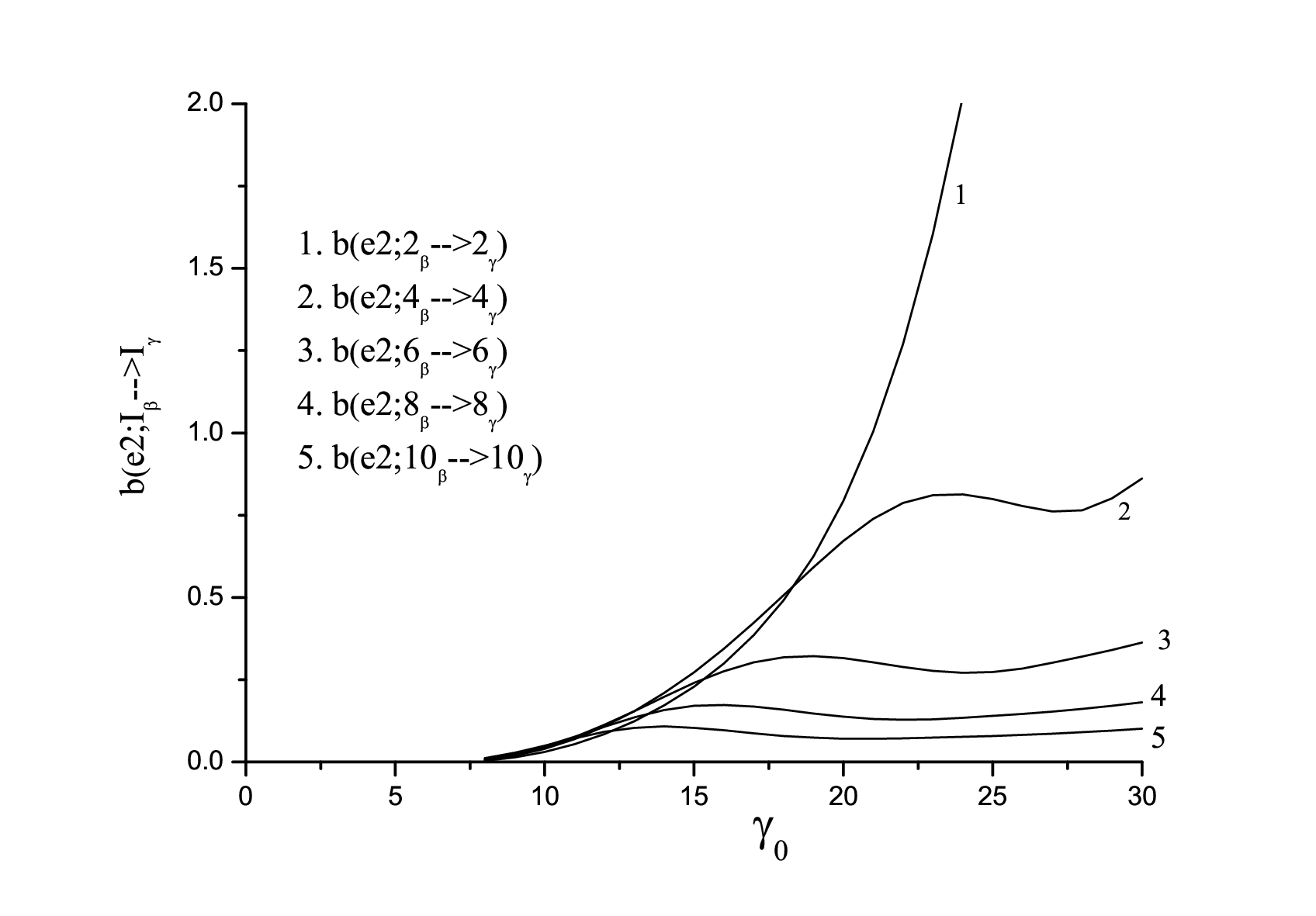}
\caption{The same as in Fig. \ref{fig1}, but for b(E2,$2I_\beta\to 2I_\gamma$).}
\label{fig9}
\end{center}
\end{figure}

\begin{figure}
\begin{center}
\includegraphics[height=8cm, width=12cm]{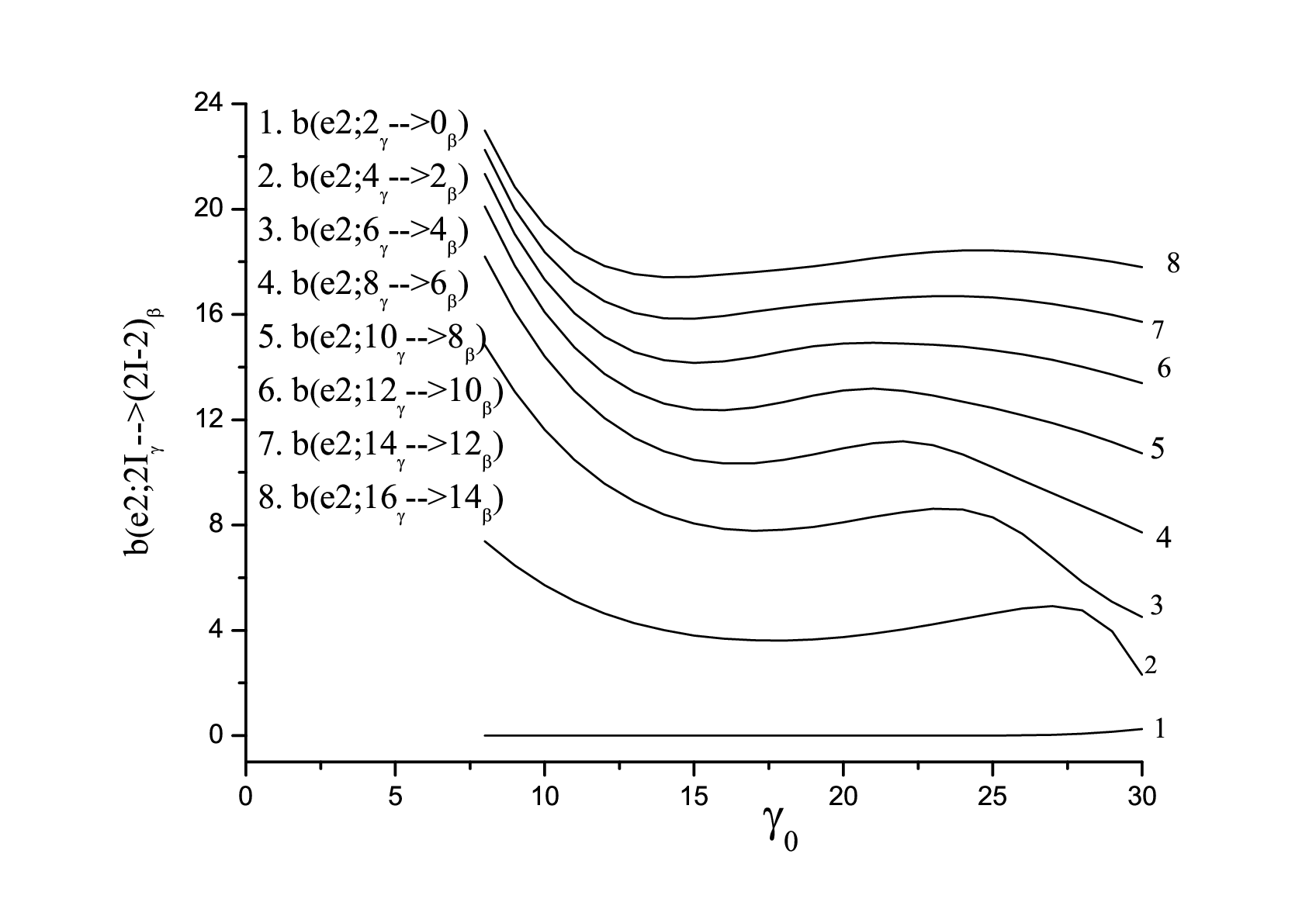}
\caption{The same as in Fig. \ref{fig1}, but for b(E2,$2I_\gamma\to (2I-2)_\beta$).}
\label{fig10}
\end{center}
\end{figure}

\begin{figure}
\begin{center}
\includegraphics[height=8cm, width=12cm]{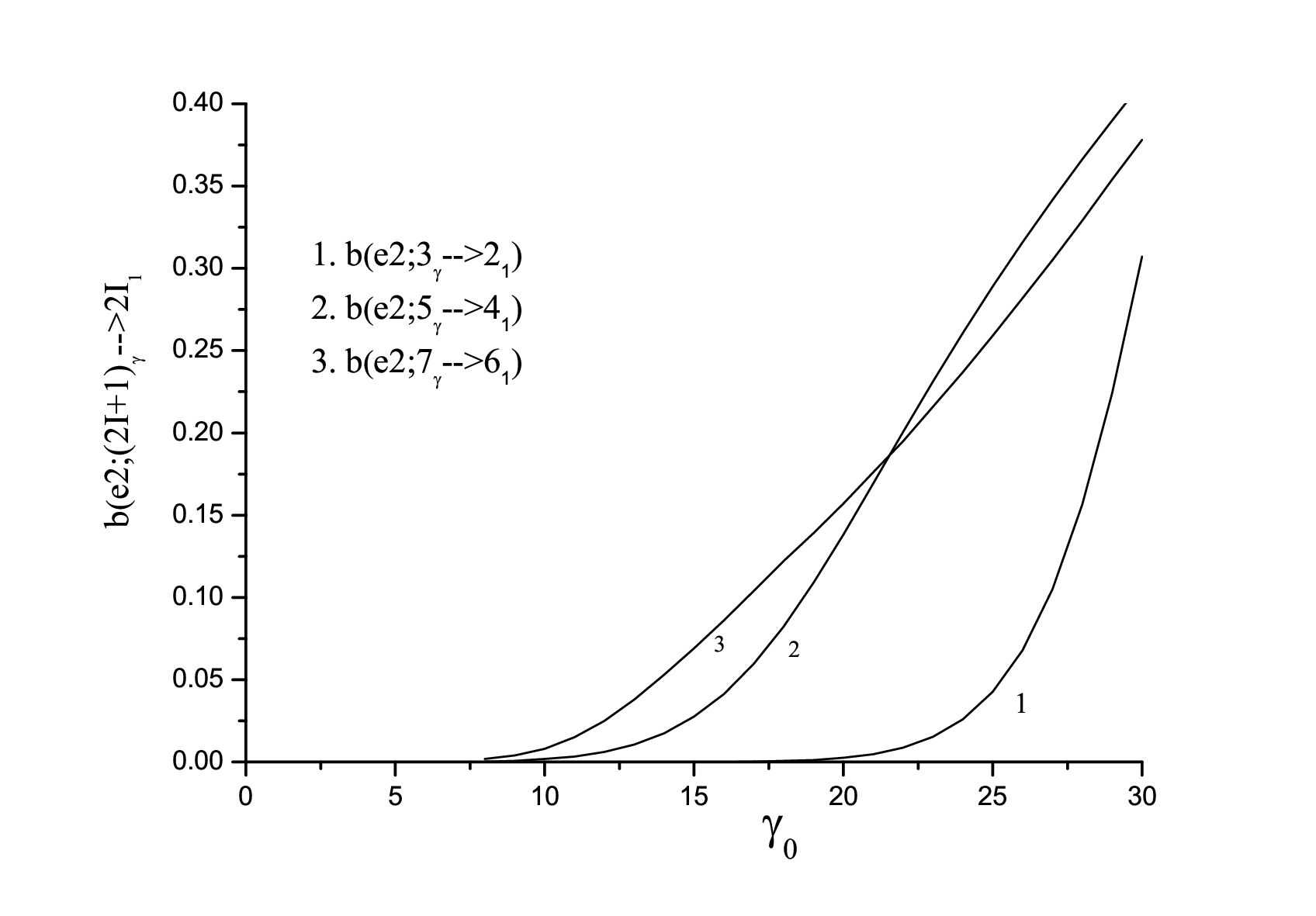}
\caption{The same as in Fig. \ref{fig1}, but for b(E2,$(2I+1)_\gamma\to 2I_1$).}
\label{fig11}
\end{center}
\end{figure}

\begin{figure}
\begin{center}
\includegraphics[height=8cm, width=12cm]{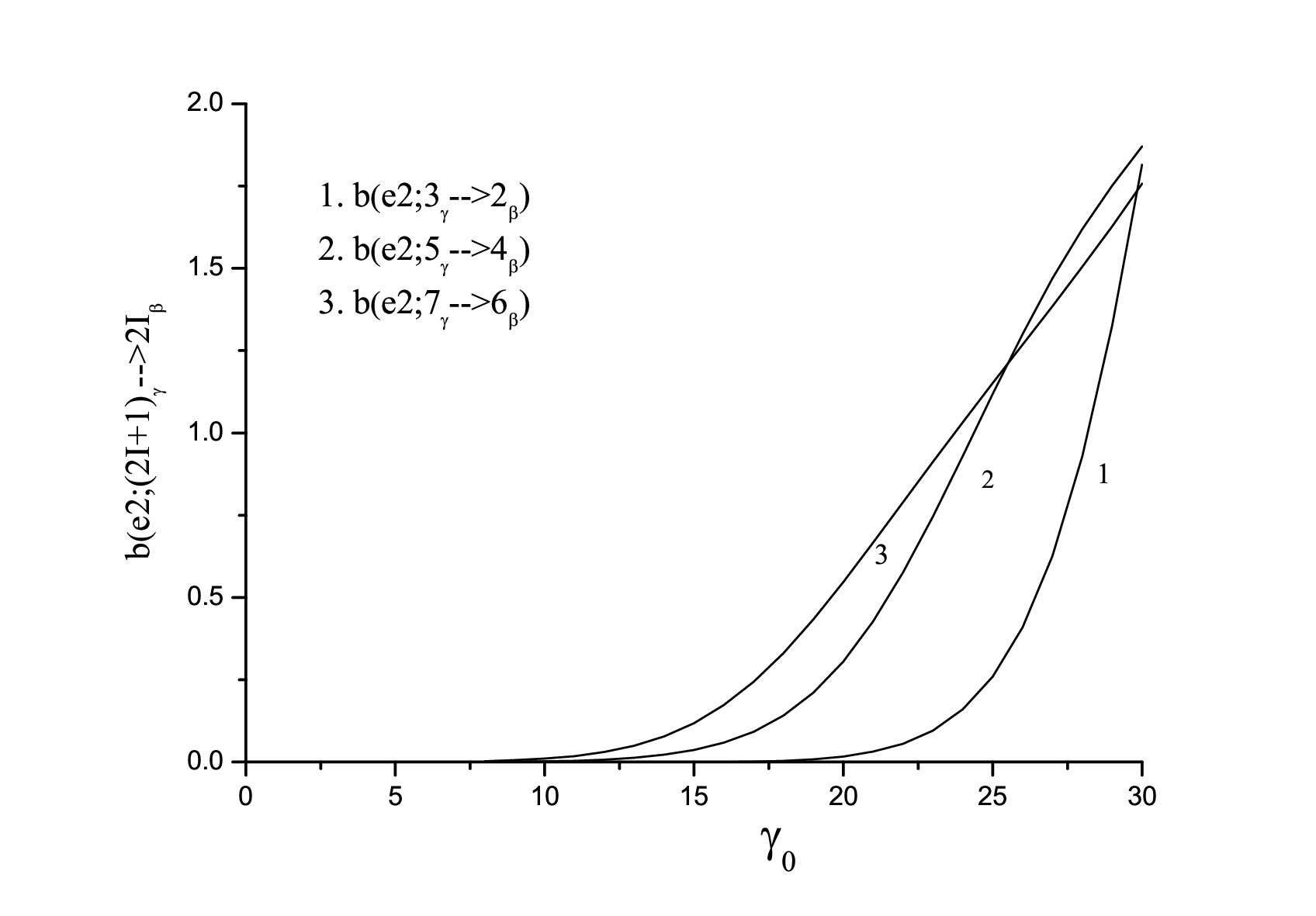}
\caption{The same as in Fig. \ref{fig1}, but for b(E2,$(2I+1)_\gamma\to 2I_\beta$).}
\label{fig12}
\end{center}
\end{figure}

In fig. \ref{fig6}-\ref{fig12} inter-band reduced \emph{E}2-transition
probabilities between ground-state-band and $\beta$-band,
ground-state-band and $\gamma$-band, $\beta$-band and $\gamma$-band
are presented.

In fig. \ref{fig6} behavior of the inter-band reduced \emph{E}2-transitions probabilities b(E2,$2I_\gamma\to (2I-2)_1$) is presented. b(E2,$2I_\gamma\to (2I-2)_1$) are more sensitivity at values $20^0\leq\gamma_0\leq 30^0$ of the  parameter $\gamma_0$. These transitions are observed in the collective excitation spectra of most heavy deformed even-even nuclei: $^{150}$Nd, $^{152,154}$Sm, $^{156,158}$Dy, $^{156}$Gd, $^{162,164,166,168}$Er, $^{188}$Yb, $^{230,232}$Th, $^{234,238}$U \cite{nndc}.

In fig. \ref{fig7} behavior of the inter-band reduced \emph{E}2-transitions probabilities between same spins $2I_\gamma\to 2I_1$ is presented. These transitions are observed in the collective excitation spectra of the following deformed even-even nuclei: $^{150}$Nd,$^{156,158}$Dy, $^{156}$Gd, $^{162,164,166,168}$Er, $^{188}$Yb, $^{230,232}$Th, $^{234,238}$U \cite{nndc}.

In fig. \ref{fig8} behavior of the inter-band reduced \emph{E}2-transitions probabilities between $2I_\beta\to (2I-2)_1$ is presented. These transitions are observed in the collective excitation spectra of the following deformed even-even nuclei: $^{150}$Nd, $^{152,154}$Sm, $^{156,158}$Dy, $^{156}$Gd, $^{162,164,166,168}$Er, $^{188}$Yb, $^{230,232}$Th, $^{234,238}$U \cite{nndc}.

In fig. \ref{fig9} behavior of the inter-band reduced \emph{E}2-transitions probabilities between same spins $2I_\beta\to 2I_\gamma$ is presented. These transitions are observed in the collective excitation spectra of the following deformed even-even nuclei: $^{152}$Sm, $^{168}$Er \cite{nndc}.

In fig. \ref{fig10} behavior of the inter-band reduced
\emph{E}2-transitions probabilities between same spins $2I_\gamma\to (2I-2)_\beta$ is presented. These transitions are observed in the collective excitation spectra of the following deformed even-even nuclei: $^{152}$Sm,$^{156}$Gd \cite{nndc}.

In fig. \ref{fig11} behavior of the inter-band reduced
\emph{E}2-transitions probabilities between same spins $(2I+1)_\gamma\to 2I_1$ is presented. These transitions are observed in the collective excitation spectra of the following deformed even-even nuclei: $^{152}$Sm,  $^{156}$Gd, $^{166,168}$Er \cite{nndc}.

In fig. \ref{fig12} behavior of the inter-band reduced
\emph{E}2-transitions probabilities between same spins $(2I+1)_\gamma\to 2I_\beta$ is presented. These transitions are observed in the collective excitation spectra of the following deformed even-even nucleus: $^{168}$Er \cite{nndc}.

The complex behavior inter/intra-band reduced \emph{E}2-transitions probabilities fig. \ref{fig1}-\ref{fig12} at large values of the $\gamma$ triaxiality parameter is ground state band and $\gamma$-band bandcrossing.

\section{Conclusion}
In this paper the triaxality dynamic in the quadrupole shape excitation in even-even heavy nuclei is considered in the framework of the free triaxiality model, where  $\gamma$ and $\beta_2$ variables are dynamic. The Davidson potential is used for $\gamma$ and $\beta_2$ variables. The energy levels of excited collective states was obtained by taking into account the high-order terms of the rotational energy operator series ($\gamma$-$\gamma_0$). The bandcrossing ground-state-band and $\gamma$-band is observed. It shows that within the framework of the free triaxiality model it is impossible to describe states with large or maximum triaxiality. If such states are detected in experiments.

The branching ratio of intra/inter-band reduced E2-transition probabilities ground-state-band, $\beta$- and $\gamma$-bands for heavy even-even nuclei are considered. The sensitivity of the intra/inter-band reduced \emph{E}2-transition probabilities to 
the triaxiality parameter $\gamma_0$ was analyzed.

\begin{table}
\begin{center}
\caption{\bf  \label{table1} Comparison of theoretical and experimental values \cite{nndc} for nuclei of rare earth elements and actinides ratio R$_{4/2}$ = E($4_1$)/E($2_1$), R$_{0_\beta}$=E(0$_\beta$)/ E($2_1$) and R$_{2_\gamma}$=E($2_\gamma$)/E($2_1$) respectively. As well as the values of the parameters $\hbar\omega$, $\mu_\beta$, $\mu_\gamma$, $\gamma_0$ and  RMS (in kEv). }
\end{center}
\begin{center}
\begin{tabular}{|c|c|c|c|c|c|c|c|c|c|c|c|}
\cline{1-12} \hline
Nuclei&R$^{exp}_{4/2}$&R$^{theor}_{4/2}$&R$^{exp}_{0_\beta}$&
 R$^{theor}_{0_\beta}$
&R$^{exp}_{2_\gamma}$&R$^{theor}_{2_\gamma}$&$\hbar\omega$&$\mu_\beta$&$\mu_\gamma$
&$\gamma_0$&RMS\\
\hline
$^{150}$Nd&2.93&3.15&5.18&6.07&8.15&9.19&369.3&0.39&1.47&12$^0$&35.6\\
$^{154}$Sm&3.25&3.28&12.36&13.16&17.58&18.66&538.6&0.27&1.36&8.9$^0$&41.6\\
$^{156}$Dy&2.93&3.05&4.9&4.81&6.46&6.57&349.4&0.44&0.89&13.9$^0$&64.9\\
$^{156}$Gd&3.24&3.27&11.8&11.73&12.98&13.45&519.5&0.28&0.21&10.4$^0$&98\\
$^{158}$Dy&3.21&3.26&10.02&10.79&9.57&10.24&520.3&0.29&0.28&12.6$^0$&75.3\\
$^{162}$Gd&3.3&3.29&19.94&19.18&12.08&11.58&712.9&0.22&0.98&11.5$^0$&13.4\\
$^{162}$Er&3.23&3.24&10.65&10.21&8.83&8.89&524.7&0.3&0.54&12.8$^0$&36.82\\
$^{164}$Er&3.28&3.25&13.63&11.93&9.41&8.86&569.8&0.28&0.2&13$^0$&85.6\\
$^{166}$Dy&3.31&3.27&15&13.31&11.19&9.91&578&0.26&0.17&12.5$^0$&42.2\\
$^{168}$Er&3.31&3.26&15.25&12.41&10.29&8.48&573.9&0.27&0.2&13.2$^0$&67.6\\
$^{168}$Yb&3.27&3.27&13.17&12.95&11.22&11.20&567.9&0.27&0.21&11.5$^0$&9.9\\
$^{168}$Hf&3.11&3.18&7.59&7.45&7.06&7.45&456.8&0.35&0.2&13.6$^0$&35.4\\
$^{170}$Hf&3.19&3.23&8.73&8.72&9.54&9.92&446.4&0.32&1.75&12.2$^0$&32.6\\
$^{170}$W&2.95&3.18&6.08&7.57&5.98&6.78&523.3 &0.34&0.&14.4$^0$&40.8\\
$^{228}$Th&3.23&3.30&14.4&15.64&16.78&18.7&409.7&0.25&0.17&8.9$^0$&14\\
$^{230}$Th&3.27&3.27&11.92&11.32&14.67&12.63&333.7&0.3&0.21&10.6$^0$&71.9\\
$^{232}$Th&3.28&3.29&14.8&13.75&15.91&14.9&356.5&0.26&0.22&9.8$^0$&87.1\\
$^{232}$U&3.29&3.29&14.53&13.46&18.22&17.76&333.4&0.26&1.2&9.2$^0$&25.7\\
$^{234}$U&3.29&3.31&18.62&19.34&21.3&22.25&408.3&0.22&1.03&8.3$^0$&12.1\\
$^{236}$U&3.3&3.31&20.31&20.47&21.17&21.66&454.2&0.22&0.21&8.4$^0$&5.5\\
$^{238}$U&3.3&3.31&20.64&19.56&23.61&24.44&427.5&0.22&0.11&7.9$^0$&44\\
$^{240}$Pu&3.31&3.3&20.1&16.43&26.55&26.04&372.9&0.24&1.55&7.8$^0$&104\\
\hline
\end{tabular}
\end{center}
\end{table}

\begin{table}
	\begin{center}
		\caption{\bf  \label{table220} The sequence of energies of the levels of the gsb- and $\gamma$- bands for different values of the nonaxiality parameter $\gamma_0$}
	\end{center}
	\begin{center}
		\begin{tabular}{|c|c|c|c|c|c|c|c|c|c|c|c|c|c|c|c|}
			\cline{1-16}
			\hline
			\multicolumn{4}{|c|}{$\gamma_0$=1$^0$}& \multicolumn{2}{c|}{$\gamma_0$=5$^0$}& \multicolumn{2}{c|}{$\gamma_0$=10$^0$}&
			\multicolumn{2}{c|}{$\gamma_0$=15$^0$}& \multicolumn{2}{c|}{$\gamma_0$=20$^0$}& \multicolumn{2}{c|}{$\gamma_0$=25$^0$}& \multicolumn{2}{c|}{$\gamma_0$=30$^0$}\\
			\hline
			I$_1$&E$_1$&I$_\gamma$&E$_\gamma$&
			E$_1$&E$_\gamma$&E$_1$&E$_\gamma$&E$_1$&E$_\gamma$&E$_1$
			&E$_\gamma$&E$_1$
			&E$_\gamma$&E$_1$
			&E$_\gamma$\\
			\hline
			2$_1$&  4& 2$_2$&6567&  4&264&  4& 67&  4& 31&  5& 18&  5& 13&  6& 12\\
			&     & 3$_1$&6571&   &268&   & 71&   & 35&   & 23&   & 19&   & 18\\
			4$_1$& 28& 4$_2$&6576& 28&274& 29& 77& 30& 42& 30& 32& 30& 31& 30& 34\\
			&     & 5$_1$&6583&   &200&   & 84&   & 49&   & 39&   & 36&   & 36\\
			6$_1$& 73& 6$_2$&6591& 74&289& 76& 93& 75& 60& 72& 54& 70& 58& 70& 60\\
			&     & 7$_1$&6600&   &298&   &103&   & 69&   & 60&   & 58&   & 58\\
			8$_1$&140& 8$_2$&6611&141&309&142&115&136& 85&130& 85&127& 89&126& 90\\
			&     & 9$_1$&6623&   &321&   &127&   & 94&   & 86&   & 84&   & 84\\
			10$_1$&228&10$_2$&6636&230&335&227&143&214&118&205&122&199&124&198&124\\
			&     &11$_1$&6651&   &349&   &156&   &125&   &117&   &114&   &114\\
			12$_1$&337&12$_2$&6667&340&366&329&177&309&158&296&164&288&163&286&162\\
			&     &13$_1$&6685&   &383&   &191&   &161&   &153&   &149&   &148\\
			14$_1$&468&14$_2$&6703&470&403&449&217&421&205&403&209&393&205&390&204\\
			&     &15$_1$&6723&   &422&   &232&   &202&   &193&   &187&   &186\\
			16$_1$&620&16$_2$&6745&622&445&587&264&551&258&527&257&514&252&510&250\\
			&     &17$_1$&6767&   &467&   &278&   &248&   &236&   &230&   &228\\
			18$_1$&793&18$_2$&6791&793&492&743&318&697&315&668&310&651&302&646&300\\
			&     &19$_1$&6817&   &517&   &329&   &298&   &284&   &276&   &274\\
			20$_1$&988&20$_2$&6843&984&546&917&378&861&377&825&366&804&357&798&354\\
			\hline
		\end{tabular}
	\end{center}
\end{table}

\begin{table}
\begin{center}
\caption{\bf The experimental values of the excited states\cite{nndc} of the $\gamma$-band of even-
even nuclei}\label{table2}
\end{center}
\begin{center}
\begin{tabular}{|c|c|c|c|c|c|c|c|c|c|c|c|}
\hline
Nucl.&$I_\gamma$&Exp.&Nucl.&$I_\gamma$&Exp.&Nucl.&$I_\gamma$&Exp.&Nucl.&$I_\gamma$&Exp.\\
\hline
$^{150}$Nd&$2_\gamma$&1061&&$3_\gamma$&1044&&$5_\gamma$&1302&&$4_\gamma$&1023\\
&$3_\gamma$&1200&&$4_\gamma$&1163&&$6_\gamma$&1445&&$5_\gamma$&1090\\
&$4_\gamma$&1353&&$5_\gamma$&1314&&$7_\gamma$&1618&&$6_\gamma$&1172\\
$^{154}$Sm&$2_\gamma$&1440&&$6_\gamma$&1486&$^{170}$W&$2_\gamma$&937&&$7_\gamma$&1261\\
&$3_\gamma$&1539&&$7_\gamma$&1675&&$3_\gamma$&1073&&$8_\gamma$&1365\\
&$4_\gamma$&1664&&$8_\gamma$&1892&&$4_\gamma$&1220&$^{236}$U&$2_\gamma$&958\\
&$5_\gamma$&1805&$^{162}$Gd&$2_\gamma$&864&$^{170}$Hf&$2_\gamma$&961&&$3_\gamma$&1001\\
&$6_\gamma$&1974&&$3_\gamma$&930&&$3_\gamma$&1087&&$4_\gamma$&1059\\
&$7_\gamma$&2154&&$4_\gamma$&1015&&$4_\gamma$&1227&&$5_\gamma$&1127\\
$^{156}$Dy&$2_\gamma$&890&$^{162}$Er&$2_\gamma$&900&$^{228}$Th&$2_\gamma$&969&$^{238}$U&$2_\gamma$&1060\\
&$3_\gamma$&1022&&$3_\gamma$&1002&&$3_\gamma$&1022&&$3_\gamma$&1105\\
&$4_\gamma$&1168&&$4_\gamma$&1128&&$4_\gamma$&1091&&$4_\gamma$&1163\\
&$5_\gamma$&1336&&$5_\gamma$&1286&&$5_\gamma$&1174&&$5_\gamma$&1232\\
&$6_\gamma$&1525&$^{164}$Er&$2_\gamma$&860&$^{230}$Th&$2_\gamma$&781&&$6_\gamma$&1311\\
&$7_\gamma$&1729&&$3_\gamma$&946&&$3_\gamma$&825&&$7_\gamma$&1403\\
&$8_\gamma$&1958&&$4_\gamma$&1058&&$4_\gamma$&883&&$8_\gamma$&1504\\
&$9_\gamma$&2191&&$5_\gamma$&1197&&$5_\gamma$&955&&$9_\gamma$&1619\\
&$10_\gamma$&2448&&$6_\gamma$&1359&&$6_\gamma$&1039&&$10_\gamma$&1741\\
&$11_\gamma$&2712&&$7_\gamma$&1545&&$7_\gamma$&1134&&$11_\gamma$&1875\\
&$12_\gamma$&2997&&$8_\gamma$&1745&&$8_\gamma$&1243&&$12_\gamma$&2018\\
&$13_\gamma$&3273&&$9_\gamma$&1977&&$9_\gamma$&1358&&$13_\gamma$&2171\\
$^{156}$Gd&$2_\gamma$&1154&&$10_\gamma$&2184&&$10_\gamma$&1520&&$14_\gamma$&2333\\
&$3_\gamma$&1248&&$11_\gamma$&2479&$^{232}$Th&$2_\gamma$&785&&$15_\gamma$&2502\\
&$4_\gamma$&1355&&$12_\gamma$&2733&&$3_\gamma$&829&&$16_\gamma$&2683\\
&$5_\gamma$&1507&&$13_\gamma$&3027&&$4_\gamma$&890&&$17_\gamma$&2868\\
&$6_\gamma$&1644&$^{166}$Dy&$2_\gamma$&857&&$5_\gamma$&960&&$18_\gamma$&3065\\
&$7_\gamma$&1850&&$3_\gamma$&928&&$6_\gamma$&1051&&$19_\gamma$&3265\\
&$8_\gamma$&2012&&$4_\gamma$&1023&&$8_\gamma$&1258&&$20_\gamma$&3474\\
&$9_\gamma$&2249&&$5_\gamma$&1141&&$9_\gamma$&1370&&$21_\gamma$&3686\\
&$10_\gamma$&2442&$^{168}$Hf&$2_\gamma$&876&&$10_\gamma$&1512&&$22_\gamma$&3906\\
&$11_\gamma$&2687&&$3_\gamma$&1031&&$11_\gamma$&1640&&$23_\gamma$&4127\\
&$12_\gamma$&2923&&$4_\gamma$&1161&&$12_\gamma$&1801&&$24_\gamma$&4358\\
&$13_\gamma$&3175&&$5_\gamma$&1386&$^{232}$U&$2_\gamma$&867&&$25_\gamma$&4586\\
&$14_\gamma$&3438&&$6_\gamma$&1551&&$3_\gamma$&912&&$26_\gamma$&4825\\
&$15_\gamma$&3715&$^{168}$Yb&$2_\gamma$&984&&$4_\gamma$&971&$^{240}$Pu&$2_\gamma$&1137\\
&$16_\gamma$&3995&&$3_\gamma$&1067&$^{234}$U&$2_\gamma$&927&&$3_\gamma$&1177\\
$^{158}$Dy&$2_\gamma$&946&&$4_\gamma$&1171&&$3_\gamma$&968&&$4_\gamma$&1232\\
\hline
\end{tabular}
\end{center}
\end{table}

\end{document}